\documentclass[a4paper,11pt]{article}
\usepackage{pos}

\title{Exploring chiral dynamics with low-energy electron- and positron-nucleon scattering}
\ShortTitle{Chiral dynamics with low-energy $e^\pm N$ scattering}

\author*{C.~Weiss}

\affiliation{Theory Center, Jefferson Lab, Newport News, VA 23606, USA}

\emailAdd{weiss@jlab.org}

\abstract{Low-energy lepton-nucleon scattering probes various expressions of the long-distance
dynamics in the baryon sector of QCD, governed by chiral symmetry breaking and the $\Delta$ resonance.
We discuss some specific studies that could be performed with combined electron- and positron-nucleon
scattering at lab energies $E_{e\pm} \lesssim$ 500 MeV. They include: (a)~chiral dynamics
in the low-$Q^2$ nucleon elastic form factors (peripheral charge and current densities);
(b)~two-photon exchange effects in the target normal
single-spin asymmetry and beam charge asymmetry in elastic and inclusive scattering ($\Delta$ excitation);
(c)~nucleon generalized polarizabilities in virtual Compton scattering with beam charge asymmetry measurements.}

\FullConference{International Workshop on Low Energy Electron Positron Physics at Jefferson Lab (LEEPP2026)\\
23--27 March 2026\\
Thomas Jefferson National Accelerator Facility, Virginia, USA\\}

\begin{document}
\maketitle

\section{Long-distance dynamics in lepton-nucleon scattering}
Lepton-nucleon scattering at energies of a few 100 MeV probes various expressions of the
effective long-distance dynamics in the baryon sector of QCD. It is governed by the
spontaneous breaking of chiral symmetry in QCD, which provides the pion as a low-mass excitation
mediating long-distance interactions; and by the $\Delta$ isobar, a low-lying nucleon resonance
with strong electromagnetic coupling. The
dynamics can be constructed and solved using methods of chiral effective field theory (EFT),
the $1/N_c$ expansion of QCD, or combinations of the two in the form of a small-scale expansion;
references are given in the following sections. The EFT methods can be combined with general techniques
of complex analyticity and dispersion relations, which connect different kinematic regions and extend the
reach of the predictions. They are synergistic with lattice QCD and can incorporate its
results as dynamical input.

Electron scattering experiments at MAMI at Mainz and CEBAF at JLab have 
performed extensive studies of long-distance dynamics in structures such as the nucleon
elastic form factors, generalized polarizabilities, resonance transition form factors, and others.
The positron beams available at MUSE at PSI \cite{MUSE:2017dod}, and projected
at JLab \cite{Accardi:2020swt} and Mainz \cite{Backe:2022unk}, create the prospect of extending this
program through combined measurements with positron and electron scattering under the same conditions.
Similar measurements with positive and negative muon scattering, and comparing electron and muon scattering,
are planned at MUSE \cite{MUSE:2017dod}. Positron scattering experiments
at higher energies have been performed by OLYMPUS at DESY \cite{OLYMPUS:2013lem}.

In this note we discuss some studies in long-distance dynamics that could be performed with
combined electron- and positron-nucleon scattering at lab energies $E_{e\pm} \lesssim$ 500 MeV.
They include: (a)~chiral dynamics in the low-$Q^2$ nucleon elastic form factors (peripheral
charge and current densities);
(b)~two-photon exchange effects in the target normal single-spin asymmetry and
beam charge asymmetry in elastic and inclusive scattering ($\Delta$ excitation);
(c)~nucleon generalized polarizabilities in virtual Compton scattering with beam charge asymmetry
measurements. We focus on the structures and dynamics and discuss the specific
contributions of positron scattering; we do not attempt to review the extensive
literature on the theory or the results in electron scattering.

The present discussion covers $e^\pm$ scattering on the nucleon.
We consider the theoretical structures for both proton and neutron; for the proton they can
be measured directly in experiments with proton targets; we do not discuss here the
questions arising in the extraction of neutron structure from experiments with nuclear targets.
Also, the discussion does not cover $e^\pm$ scattering on nuclei ($A > 1$)
for measurements of nuclear form factors, radii, two-photon exchange effects
in elastic scattering, and related structures. Such processes involve
nuclear structure and nuclear excitations in the 1-10 MeV energy range
and require separate discussion.

\section{Chiral dynamics in low-$Q^2$ elastic form factors}
\label{sec:elastic}

Elastic scattering $e^\pm p \rightarrow e^\pm p$ is used to measure the proton
electric and magnetic form factors, $G_{E,M}(t)$, which are the most basic expression of the
extended spatial structure of the nucleon. A particular characteristic are the
electric and magnetic radii, $\langle r^2 \rangle_{E, M} \equiv 6 G'_{E, M}(0) / G_{E, M}(0)$;
for their interpretation in a relativistic context, see \cite{Miller:2018ybm}.
Much more information is contained in the functional dependence of the form factors
at $Q^2 \equiv -t > 0$, especially at $Q^2 \sim M_\pi^2 \sim \textrm{few}\, 10^{-2} \, \textrm{GeV}^2$.

Long-distance dynamics is expressed in the complex-analytic structure of the form factors.
The $\pi\pi$ cut at $t > 4 M_\pi^2$
results from $t$-channel processes in which the current couples to the proton through
exchange of a $\pi\pi$ state. This singularity governs the peripheral charge and current
densities in the nucleon; for a rigorous formulation in terms of transverse densities, see
\cite{Strikman:2010pu,Granados:2013moa,Alarcon:2017asr}.
Its strength can be calculated combining unitarity in the $\pi\pi$ channel and dynamical
input from chiral EFT \cite{Alarcon:2017ivh,Leupold:2017ngs,Alarcon:2018irp,Alarcon:2022adi}; for other methods
such as dispersion theory see \cite{Hoferichter:2016duk} and references in the quoted works.
The $\pi\pi$ cut lies in the unphysical region and cannot be reached directly in scattering
experiments. However, its features are encoded in the low-$Q^2$ behavior of the spacelike
form factor, specifically in the size of higher derivatives at $Q^2 = 0$ (often expressed
as ``higher moments,'' $\langle r^{2n} \rangle$ with $n > 1$, generalizing the electric/magnetic
radii) \cite{Alarcon:2017ivh}. Precise experimental information on the $Q^2$ dependence 
of the form factor at $Q^2 \sim \textrm{few}\, 10^{-2} \, \textrm{GeV}^2$
could thus directly test the long-distance dynamics in the $\pi\pi$ cut
responsible for peripheral nucleon structure. The higher derivatives of the form factor
also control the systematic uncertainties in the proton radius extraction.

The determination of the elastic form factors from $ep$ scattering data is limited by
two-photon exchange (TPE) corrections;
see \cite{Arrington:2011dn} for a review. Measurements with positrons and electrons could be used
to isolate the TPE correction and determine its size by taking the difference $e^+ - e^-$, or to
eliminate the TPE effects by taking the sum $e^+ + e^-$.
This would be a major advance in precision measurements of low-$Q^2$ form factors.
It would not only improve the proton radius extraction but enable determination of the
higher derivatives and multiscale features of the $Q^2$ dependence for novel
studies of chiral dynamics and peripheral nucleon structure.

%
% FIGURE
%
\begin{figure}[t]
\includegraphics[width=1.0\textwidth]{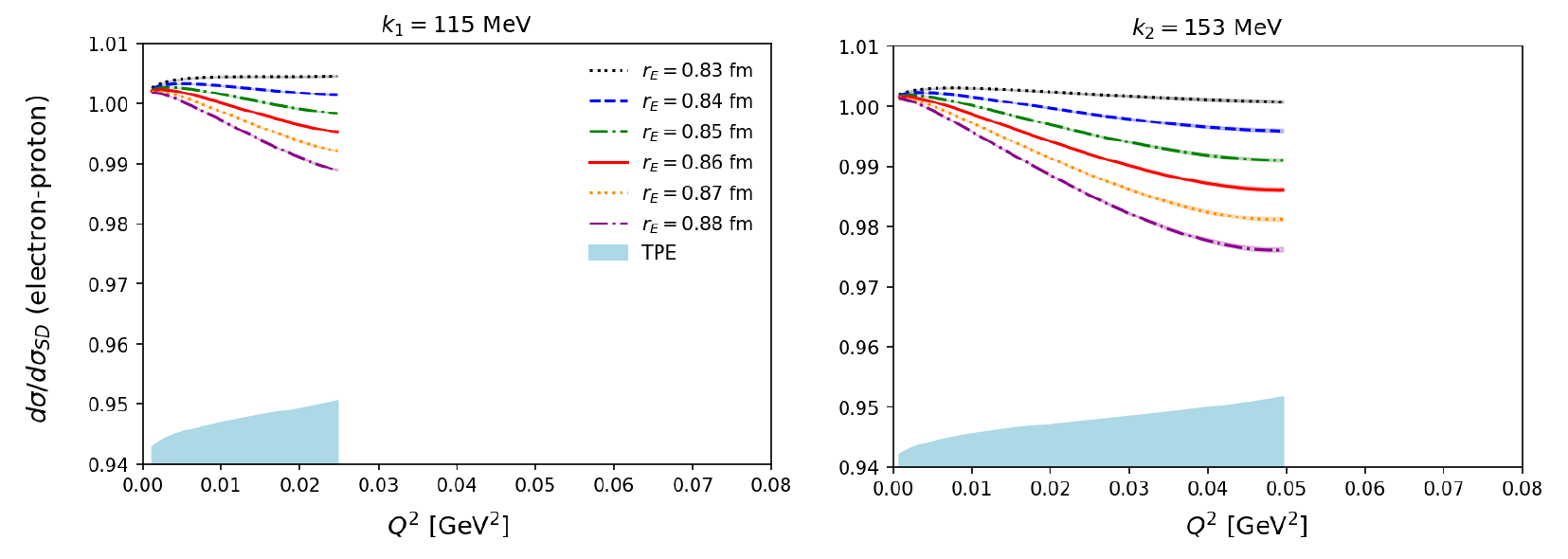}
\caption{The $Q^2$ dependence of the cross section of elastic $ep$ scattering (normalized to the
standard dipole),
for two values of the lab beam momentum $k_1$. Curves: $Q^2$ dependence predicted for a given
assumed value of the proton radius (DI$\chi$EFT) \cite{Gil-Dominguez:2023hku}.
Band at bottom: Size of TPE corrections \cite{Tomalak:2015aoa}.}
\label{fig:muse}
\end{figure}
Figure~\ref{fig:muse} shows a theoretical calculation comparing the size of the TPE correction
with the predicted radius dependence of the $ep$ elastic cross section,
in the energy range of the MUSE experiment \cite{Gil-Dominguez:2023hku}.
In the dispersively improved chiral EFT (DI$\chi$EFT) approach,
the proton radius is treated as a parameter, and the theory predicts the nonlinear
$Q^2$-dependence of the form factor for a given value of the radius.
One observes that TPE is the dominant theoretical uncertainty,
especially near the upper end of the $Q^2$ range accessible at the given energy.
Positron measurements in this energy range would thus have a major impact on the
overall determination of the form factors.

\section{Two-photon exchange in single-spin asymmetries}
\begin{figure}[t]
\includegraphics[width=1.0\textwidth]{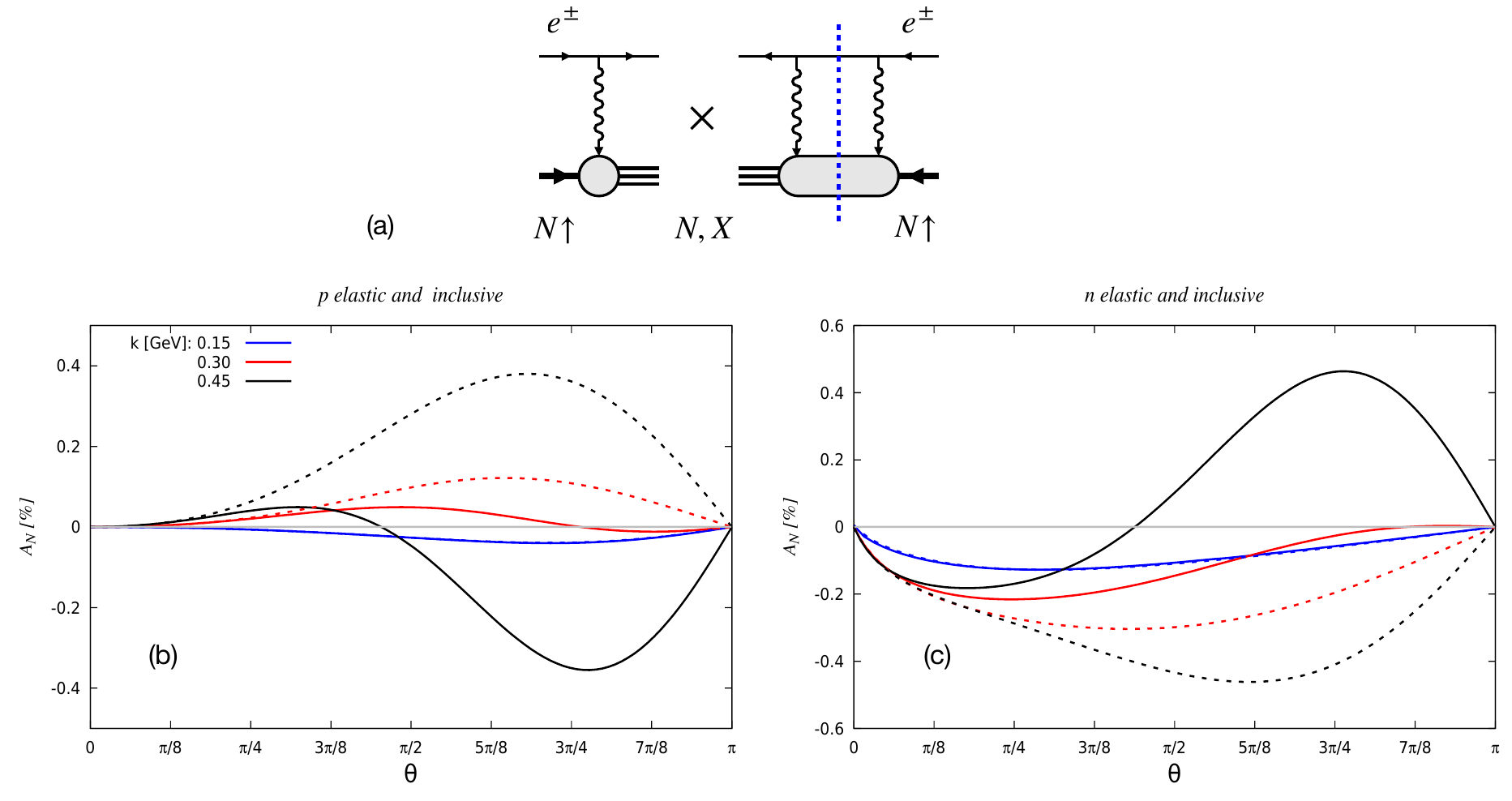}
\caption{(a) The target normal SSA $A_N$ in low-energy $e^\pm N(\uparrow)$ scattering.
(b, c) $A_N$ in $ep$ and $en$ scattering, as a function of
the CM scattering angle $\theta$, for several values of the initial electron CM momentum $k$ \cite{Goity:2023sph}.
Dashed lines: Elastic scattering. Solid lines: Inclusive scattering (includes $N$ and $\Delta$ final states)
}
\label{fig:ssa}
\end{figure}

Single-spin asymmetries (SSAs) in $e^\pm N$ scattering enable direct studies of TPE effects.
The target normal SSA in $e^\pm N(\uparrow)$ scattering is defined as
\begin{align}
A_N \equiv \frac{\sigma\uparrow - \sigma\downarrow}{\sigma\uparrow + \sigma\downarrow},
\label{ssa}
\end{align}
where the polarization is normal to the scattering plane. It can be measured in elastic
or inclusive scattering, $e N(\uparrow) \rightarrow e'N$ or $ \rightarrow e'X$.
$A_N$ is zero in one-photon exchange approximation and arises from the interference of one- and two-photon
exchange amplitudes (see Fig.~\ref{fig:ssa}a).
It involves only the imaginary part of the TPE amplitude, which is infrared-finite
and can be calculated and discussed separately from real photon emission processes.

The target normal SSA in low-energy $e^\pm N$ scattering has been analyzed
in the combined chiral and $1/N_c$ expansion \cite{Goity:2022yro,Goity:2023sph}.
This formulation of long-distance dynamics permits a systematic treatment of $\Delta$ excitation
in the intermediate and final states of the process. Figure~\ref{fig:ssa}b and c shows $A_N$ for
electron scattering on the proton and neutron, for elastic and inclusive scattering.
One observes that the asymmetries
are of the order $\sim \textrm{few} \, 10^{-3}$ and exhibit a rich kinematic dependence.
$A_N$ in elastic scattering has also been computed using empirical amplitudes \cite{Ahmed:2023zli}.
%
% FIGURE
%

As a pure TPE effect, the SSA is odd in the beam charge, $A_N(e^+N) = -A_N(e^-N)$.
Measurements with positrons could test this universality relation and validate the EFT calculations.
This would improve the theoretical understanding of TPE also in other low-energy processes such
as elastic scattering (see Sec.~\ref{sec:elastic}). Measurements of $A_N(e^+N)$ can be performed
with unpolarized positrons and appear realistic at few 100 MeV energies. Measurements in elastic
scattering at CEBAF energies (2.2-6.6 GeV) have been simulated \cite{Grauvogel:2021btg}. Measurements
in inclusive deep-inelastic scattering have been performed at HERMES \cite{HERMES:2009hsi};
for discussion of the energy/momentum dependence and parton-hadron duality in $A_N$,
see \cite{Afanasev:2007ii}.

The beam normal SSA $B_N$ in $e^\pm (\uparrow) N$ scattering is defined as in Eq.~(\ref{ssa})
but with the lepton polarized normally to the scattering plane. It is proportional to the lepton mass
and therefore small; this is partly compensated by a logarithmic enhancement due to collinear photon
exchange in the TPE amplitude \cite{Afanasev:2004pu,Borisyuk:2005rj}.
$B_N$ is estimated to be of the order $\sim 10^{-5}$ in the low-energy region discussed
here \cite{Borisyuk:2005rj}.
Measurements with positrons would require a transversely polarized positron beam
and appear challenging. $B_N$ is much larger in $\mu^\pm (\uparrow) N$ scattering \cite{Koshchii:2019mgv},
and measurements with muons appear more promising.

\section{Polarizabilities in virtual Compton scattering}
\begin{figure}[t]
\includegraphics[width=.75\textwidth]{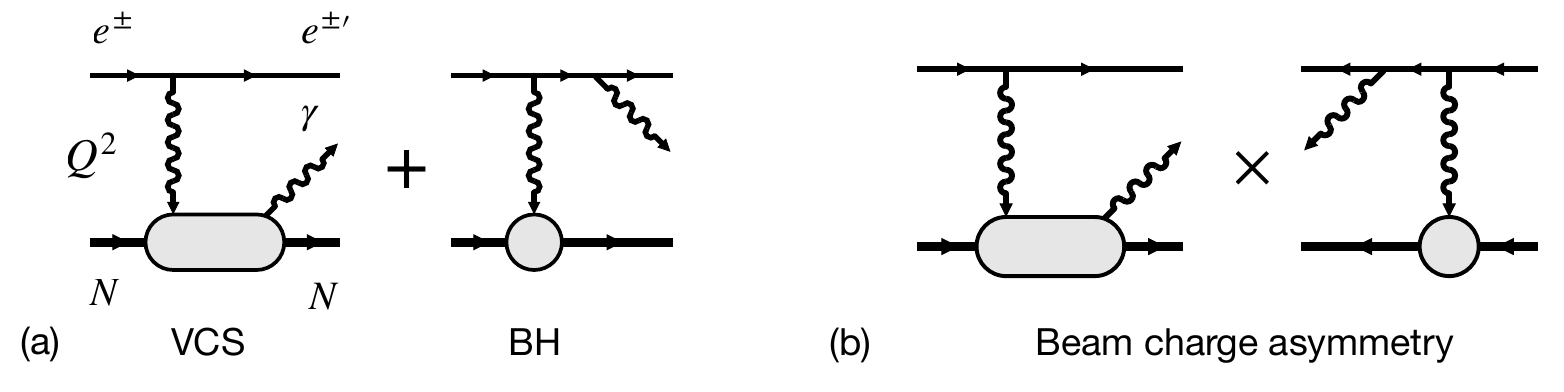}
\caption{(a) Virtual Compton Scattering and Bethe-Heitler amplitudes
in $e^\pm N \rightarrow e^{\pm \prime} \gamma N$.
(b)~Beam charge asymmetry from BH-VCS interference.}
\label{fig:vcs}
\end{figure}

Real photon production in lepton-nucleon scattering, $e^\pm N \rightarrow e^\pm \gamma N$, is the
most basic inelastic scattering process and an essential tool for exploring structure and dynamics.
The production involves the virtual Compton scattering (VCS) and Bethe-Heitler (BH) processes
(see Fig.~\ref{fig:vcs}a).
Their amplitudes interfere, giving rise to a rich structure of the cross section and observables.
%
% FIGURE
%

Long-distance dynamics in VCS has been analyzed using dispersion relations
\cite{Drechsel:2002ar,Schumacher:2005an} and chiral EFT calculations
\cite{Bernard:2007zu,Griesshammer:2012we}; see the reviews quoted
here for the extensive literature. The low-energy expansion of the VCS amplitude defines the so-called
generalized polarizabilities of the nucleon, $\alpha_E(Q^2)$ and $\beta_M(Q^2)$. They
describe the response of the charge and magnetization distributions in the hadron to
an external electromagnetic field and extend the information provided by the elastic form factors.
Experimental results from JLab and Mainz are reviewed in \cite{Fonvieille:2019eyf,Sparveris:2025vxu}.

Measurements of VCS in low-energy positron-nucleon scattering can determine the beam charge asymmetry (BCA)
of the unpolarized cross section (differential in the final-state variables),
\begin{align}
A^C_{UU} \equiv \frac{\sigma_{UU}(e^+) - \sigma_{UU}(e^-)}{\sigma_{UU}(e^+) + \sigma_{UU}(e^-)}.
\end{align}
The BCA arises from the interference of the BH and VCS amplitudes (see Fig.~\ref{fig:vcs}b)
and gives access to the real part of the VCS amplitude.
In the analysis using dispersion relations, this information constrains the subtraction constant
in the dispersion relation for the real part of the VCS amplitude. 
In the chiral EFT calculations, it determines the low-energy constant describing a local $\gamma\gamma NN$
coupling (induced by high-energy degrees of freedom which have been integrated out in the EFT).
In either approach this would improve the extraction of the generalized polarizabilities and
the knowledge of long-distance nucleon structure.
Measurements of the BCA in positron-nucleon scattering at higher energies ($\gtrsim 1$ GeV)
\cite{Sparveris:2025vxu} could
cross-check puzzling results for the generalized polarizabilities reported by the JLab VCS I
experiment \cite{VCS-II:2023jhu}, which show a non-uniform $Q^2$-dependence in conflict with
basic notions of complex analyticity \cite{Li:2022sqg}.

\section{Summary}
Experiments in combined electron- and positron-nucleon scattering at lab energies
$E_{e\pm} \lesssim$ 500 MeV can provide important and unique information for the study of
long-distance dynamics in the baryon sector of QCD, by determining the TPE effects in elastic
scattering and hadron production ($\Delta, \pi N$) and providing charge asymmetries in VCS.
Simulations of such measurements
should be performed as part of the planning for the future positron facility at JLab \cite{Accardi:2020swt}.

\vspace{1ex}
This material is based upon work supported by the U.S.~Department of Energy, Office of Science,
Office of Nuclear Physics under 
Contract No.\ 89243126CSC000213.
\end{document}